\documentclass[sigconf]{acmart}
\usepackage{textcomp}
\usepackage{algorithm}
\usepackage{wasysym}
\usepackage{algorithmic}
\usepackage{color}
\usepackage{amsmath}
\usepackage{graphicx}
\usepackage{caption}
\usepackage{subcaption}

\AtBeginDocument{%
  \providecommand\BibTeX{{%
    \normalfont B\kern-0.5em{\scshape i\kern-0.25em b}\kern-0.8em\TeX}}}

\setcopyright{acmlicensed}
\copyrightyear{2018}
\acmYear{2018}
\acmDOI{XXXXXXX.XXXXXXX}

\acmConference[Conference acronym 'XX]{Make sure to enter the correct
  conference title from your rights confirmation emai}{June 03--05,
  2018}{Woodstock, NY}
\acmISBN{978-1-4503-XXXX-X/18/06}

\begin{document}

\title[LAST: Learning At Serving Time]{A re-ranking model with pre-feedback learning at serving time in e-commerce}
\title[LAST: Learning At Serving Time]{Do Not Wait: Learning Re-Ranking Model Without User Feedback At Serving Time in E-Commerce}



\author{Yuan Wang}
\authornote{Both authors contributed equally to this research.}
\authornote{First corresponding author: Xiao Zhang (zhangx89@ruc.edu.cn),\\ second corresponding author: Yuan Wang (wy175696@alibaba-inc.com)}
\email{wy175696@alibaba-inc.com}
\author{Zhiyu Li}
\authornotemark[1]
\email{tuanyu.lzy@alibaba-inc.com}
\affiliation{%
  \institution{Alibaba Group}
  \city{Hangzhou}
  \state{Zhejiang}
  \country{China}
}

\author{Changshuo Zhang}
\affiliation{%
  \institution{Gaoling School of Artificial Intelligence, Renmin University of China}
  \city{Beijing}
  \country{China}}
\email{lyingcs@ruc.edu.cn}

\author{Sirui Chen}
\affiliation{%
  \institution{School of Information, Renmin University of China}
  \city{Beijing}
  \country{China}}
\email{chensr16@gmail.com}

\author{Xiao Zhang}
\authornotemark[2]
\affiliation{%
  \institution{Gaoling School of Artificial Intelligence, Renmin University of China}
  \city{Beijing}
  \country{China}}
\email{zhangx89@ruc.edu.cn}

\author{Jun Xu}
\affiliation{%
  \institution{Gaoling School of Artificial Intelligence, Renmin University of China}
  \city{Beijing}
  \country{China}}
\email{junxu@ruc.edu.cn}

\author{Quan Lin}
\affiliation{%
  \institution{Alibaba Group}
  \city{Hangzhou}
  \state{Zhejiang}
  \country{China}}
\email{tieyi.lq@alibaba-inc.com}

\renewcommand{\shortauthors}{Wang and Li, et al.}

\begin{abstract}
Recommender systems have been widely used in e-commerce, and re-ranking models are playing an increasingly significant role in the domain, which leverages the inter-item influence and determines the final recommendation lists. Online learning methods keep updating a deployed model with the latest available samples to capture the shifting of the underlying data distribution in e-commerce. However, they depend on the availability of real user feedback, which may be delayed by hours or even days, such as item purchases, leading to a lag in model enhancement. 
In this paper, we propose a novel extension of online learning methods for re-ranking modeling, which we term LAST, an acronym for Learning At Serving Time. It circumvents the requirement of user feedback by using a surrogate model to provide the instructional signal needed to steer model improvement. Upon receiving an online request, LAST finds and applies a model modification on the fly before generating a recommendation result for the request. The modification is request-specific and transient. It means the modification is tailored to and only to the current request to capture the specific context of the request. After a request, the modification is discarded, which helps to prevent error propagation and stabilizes the online learning procedure since the predictions of the surrogate model may be inaccurate. Most importantly, as a complement to feedback-based online learning methods, LAST can be seamlessly integrated into existing online learning systems to create a more adaptive and responsive recommendation experience. Comprehensive experiments, both offline and online, affirm that LAST outperforms state-of-the-art re-ranking models.

\end{abstract}

\ccsdesc[500]{Theory of computation~Online algorithms}
\ccsdesc[500]{Computing methodologies~Learning paradigms}
\ccsdesc[500]{Applied computing~Online shopping}

\keywords{online learning, re-ranking, surrogate model, recommender system, e-commerce}



\maketitle


\section{Introduction}
\begin{figure}
    \includegraphics[width=0.35\textwidth]{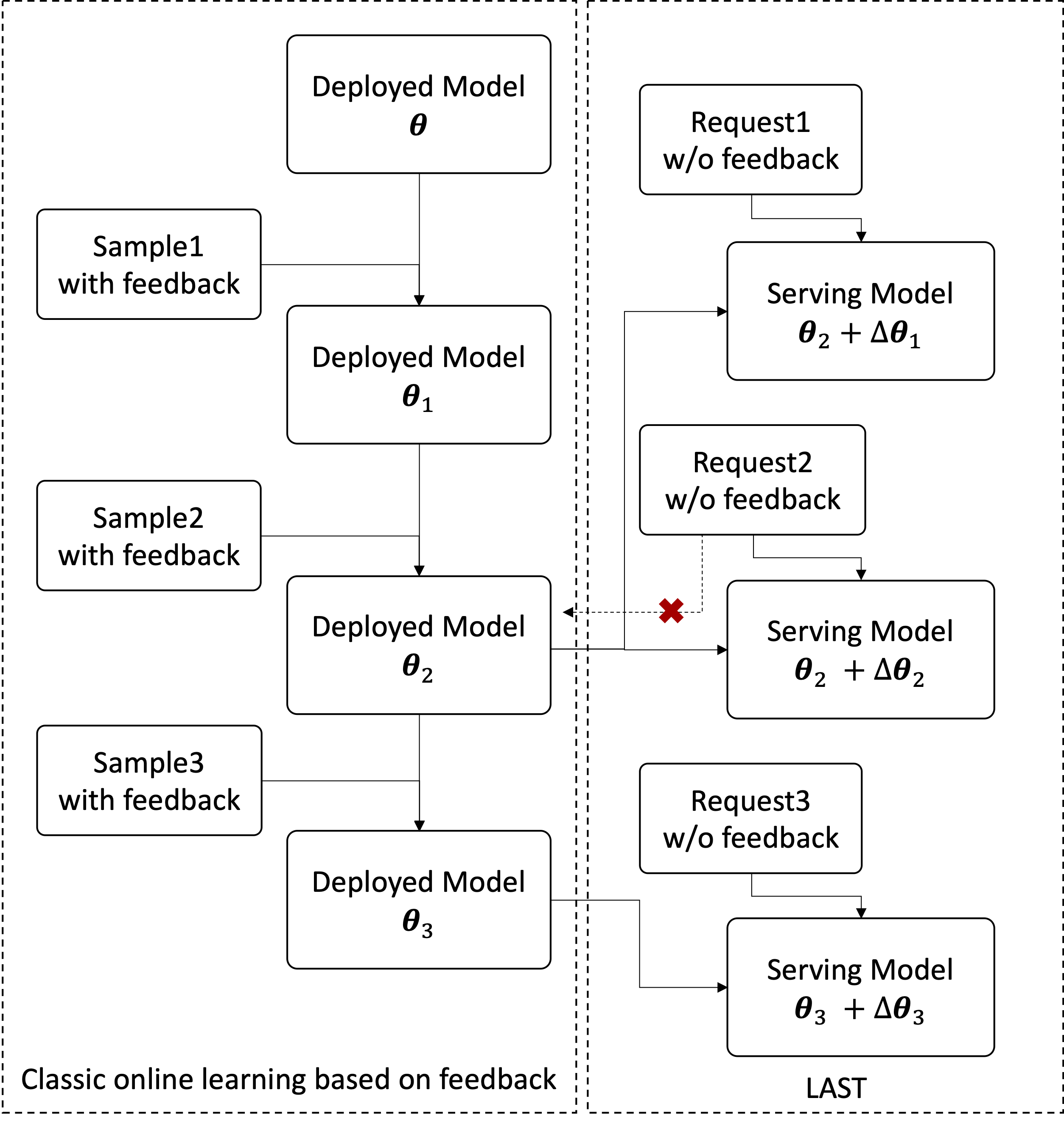}
    \caption{Classic online learning methods and our new proposal, LAST. $\boldsymbol{\theta}$ means the parameter of a model. The classic methods rely on authentic user feedback, providing enduring non-request-specific updates. LAST provides transient request-specific updates with the help of a surrogate evaluation model.  The two can work synergistically to create a more adaptive and responsive online serving system.
}
    \label{fig:online_models}
\end{figure}

Machine learning-based and deep learning-based recommendation models have become an integral part of e-commerce platforms, such as Taobao and Amazon. Re-ranking models \cite{zhuang2018globally, pei2019personalized, ai2018learning, ai2019learning, bello2018seq2slate} typically reside in the last stage of an industrial recommendation pipeline and directly determine the final recommendation lists. They explicitly consider the mutual influence between items and explore all permutations of candidates, which makes it challenging to train a re-ranking model. Reinforcement Learning (RL) \cite{kaelbling1996reinforcement, li2017deep, wiering2012reinforcement} searching algorithms and an Actor-Evaluator (AE) framework \cite{gong2019exact, huzhang2021aliexpress, wang2019sequential} have been proposed to automatically find the best recommendation list-generating policy, removing the burden of manually specifying the best recommendation list as the label in Supervised Learning (SL). After the deployment of a re-ranking model, online learning methods \cite{Sahoo2018Online, Hazan2016Introduction, Zhang2019Survey} can be applied to continuously update the deployed model using the latest available samples so that the model can capture real-time changes in the data distribution. As shown in Fig.~\ref{fig:online_models} on the left-hand side, these updates are directly integrated into the deployed model, accumulating over time and shaping all subsequent predictions. One fundamental limitation of these methods is that they depend on the availability of real user feedback, which, akin to product purchases in e-commerce, may come several hours or even days later and ultimately constrains the temporal effectiveness of the model. Moreover, at any given moment, all requests are served with a fixed model instance, lacking contextual adaptations. In a large e-commerce platform, a recommender system may receive tens of thousands of requests within a second. These requests have their own context, reflecting diverse user preferences. A single model may not be able to capture all the variety. 

In this paper, we propose LAST, an acronym for Learning At Serving Time.  It ensures continuous model optimization and fine-grained model adaptation even in situations where feedback is unattainable, as illustrated on the right-hand side in Fig.~\ref{fig:online_models}. It uses a surrogate evaluation model as the instructional signal to steer model refinement to ensure model freshness.  LAST generates transient, contextually tailored model adjustments for each request, meticulously engineered to optimize the recommendation efficacy of each request. After each recommendation, the according modification is discarded, leaving no residual influence on the deployed model. This design helps to prevent error propagation and stabilize the online learning procedure since the predictions of the surrogate model may be inaccurate. It is also more friendly to the online engineering system, as the modification functionality can be implemented as a normal model module, requiring no upgrading of the online engineering system, with or without the support of classic online learning. Most importantly, this design ensures seamless integration of LAST with existing feedback-based online learning methods. Comprehensive experiments, both offline and online, affirm that LAST outperforms state-of-the-art re-ranking models.

The main contributions of the paper are:\\
\textbf{(1) A new re-ranking model with unique online learning ability.} LAST boosts recommendation quality by feedback-independent, transient, request-specific, engineering-friendly online model modifications. It addresses the inherent temporal and adaptation limitations of conventional online learning methods and can be combined with them to enhance user satisfaction and system outcomes. \\
\textbf{(2) Comprehensive evaluations of the new proposal.} We demonstrate the effectiveness of LAST offline with publicly available data and online in an industrial environment. We have released the experimental code to increase reproducibility. 

\section{Preliminary}\label{sec:preliminary}
In this section, we formally introduce the problem definition of re-ranking modeling and the AE framework. Given a set of $M$ candidate items $\mathcal{C}=\{c_i\}_{1\leq i \leq M }$, a user $u\in \mathcal{U}$, and a list reward function $R(\cdot)$, the goal of a re-ranking model is to find the optimal list $L_{\mathcal{C}}^*$ composed by items in $\mathcal{C}$:
\begin{equation*}
L_{\mathcal{C}}^*=\underset{L_{\mathcal{C}}}{\arg\max}~ R(u, L_{\mathcal{C}}, y_{L_{\mathcal{C}}}),
\end{equation*}
where each list is of length $N$ and it is obvious that $N\leq M$.
$y_{L_{\mathcal{C}}}$ is the feedback of the user $u$ to the list $L_{\mathcal{C}}$. In e-commerce, $y$ usually involves user engagements, such as click and purchase.
$R(\cdot)$ may also consider other desirable aspects of the recommended list, such as diversity \cite{jiang2017learning, liu2020dvgan, xia2017adapting}, novelty \cite{xia2016modeling, kotkov2016survey}, and fairness \cite{karako2018using, oosterhuis2021computationally, zhu2021fairness}. 
Here, we assume the existence of a single optimal list. For simplicity, we henceforth drop the subscript $\mathcal{C}$. 
A list-generating model $G(\cdot)$, parameterized by $\boldsymbol{\theta}$, is trained to find the optimal list in a single forward execution:
\begin{equation*}
\widehat{L^*}=G(u, \mathcal{C}; \boldsymbol{\theta})
\end{equation*}
In the actor-evaluator framework, $G(\cdot)$ is called the actor. The training of the actor can be described as:
\begin{equation*}
\underset{\boldsymbol{\theta}}{\max}~ 
\mathbb{E}_{u,\mathcal{C}}[R(u, G(u, \mathcal{C}; \boldsymbol{\theta}), y_{G(u, \mathcal{C}; \boldsymbol{\theta})})],
\end{equation*}
where $\mathbb{E}$ means expectation. 
However, in the training process, the actor unavoidably generates lists that have never been shown to the user, and thus $y$ is not available. To overcome this problem, a surrogate model  $E(\cdot)$,  is trained to approximate $R(\cdot)$ in the actor-evaluator framework, namely $E(u, L; \boldsymbol{\phi}) \to R(u, L, y_L)$, where $\boldsymbol{\phi}$ is the parameter of $E(\cdot)$. The surrogate model is called the evaluator, and it is trained before the actor as 
\begin{equation*}
\underset{\boldsymbol{\phi}}{\min}~\mathbb{E}_{u,L}[diff(E(u, L; \boldsymbol{\phi}), R(u, L, y_L))].
\end{equation*}
Notably, the evaluator learns how to predict user feedback during the training process. 
Then, the training of the actor becomes:
\begin{equation*}
\underset{\boldsymbol{\theta}}{\max}~ \mathbb{E}_{u,\mathcal{C}}[E(u, G(u, \mathcal{C}; \boldsymbol{\theta}))].
\end{equation*}
It is crucial to see that the whole offline training process of the actor does not directly depend on user feedback, given the evaluator $E(\cdot)$. 

\section{LAST: The Proposed Method}\label{sec:our_method}
\begin{figure}
    \includegraphics[width=0.48\textwidth]{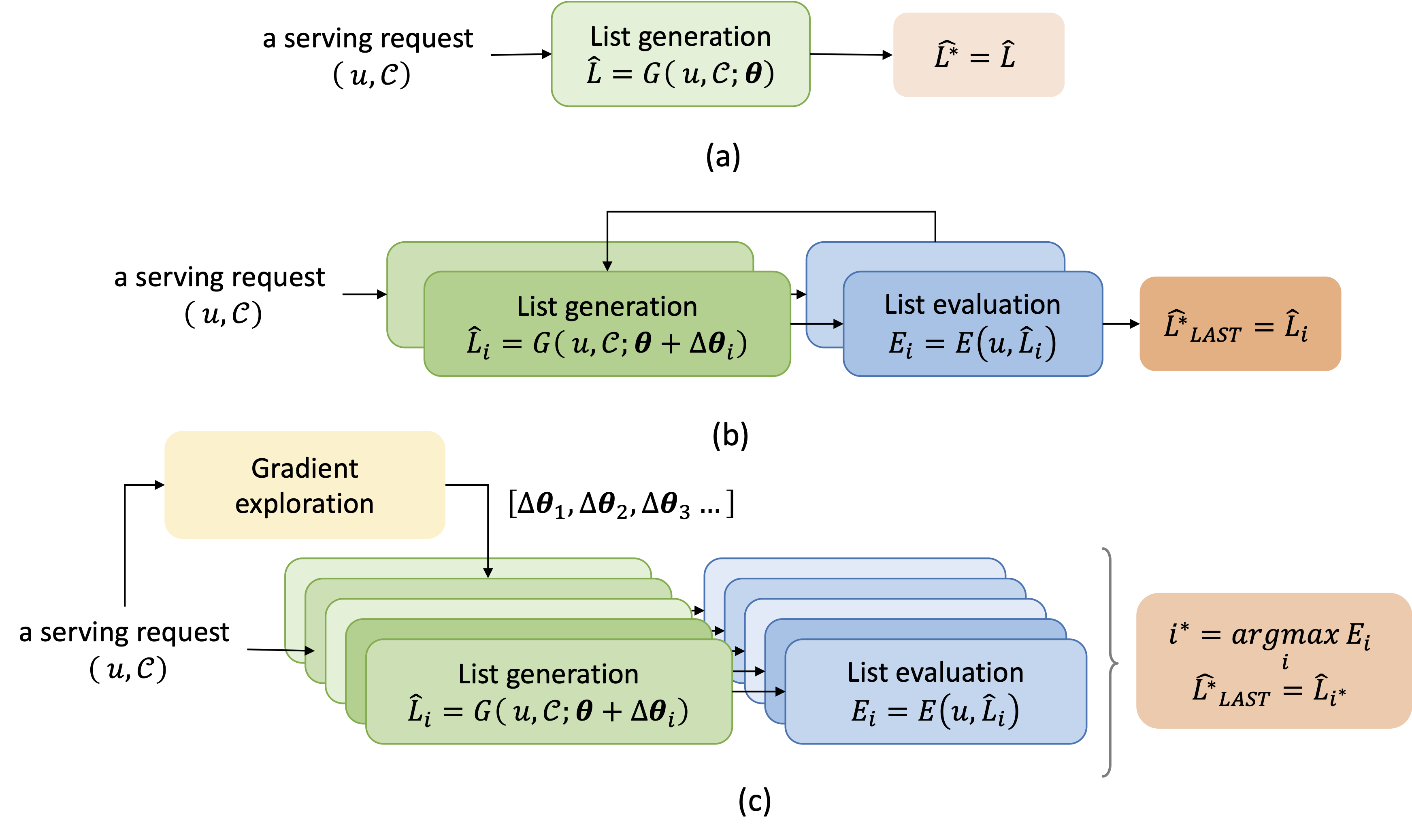}
    \caption{Online serving processes of re-ranking models. (a)~The traditional re-ranking models. The model generates a recommendation list based on its fixed policy and presents the list directly to the user. (b)~The cascade version of LAST. The actor interacts with the evaluator iteratively to improve its list-generating policy for higher evaluations. The list generated in the last iteration is presented to the user.  (c)~The parallel version of LAST. A separate gradient exploration module suggests potential model modifications. The actor tries out the suggestions, and the evaluator estimates their quality. The list with the highest evaluation is presented to the user.}
    \label{fig:LAST}
\end{figure}

Classic online learning methods rely on the real user feedback, which can be delayed by hours or even days, such as item purchases. LAST ensures continuous model optimization and fine-grained model adaptation even in the absence of user feedback. When a user request, denoted as $(u, \mathcal{C})$, is received, LAST adds a request-specific disposable modification, $\Delta \boldsymbol{\theta}$, to enhance the deployed model:
\begin{equation*}
\widehat{L^*}_{LAST}=G(u, \mathcal{C}; \boldsymbol{\theta}+ \Delta \boldsymbol{\theta}^*(u, \mathcal{C})).
\end{equation*}
The optimal modification $\Delta \boldsymbol{\theta}^*(u, \mathcal{C})$ can be obtained with the help of the evaluator:
\begin{equation*}
\label{eq:optimal_model_modifycation}
\Delta \boldsymbol{\theta}^*(u, \mathcal{C}) = \underset{\Delta \boldsymbol{\theta}}{\arg\max}~ E(u, G(u, \mathcal{C};\boldsymbol{\theta} + \Delta \boldsymbol{\theta})).
\end{equation*}
This modification is computed on the fly, tailored to the individual context of the request, capturing the unique needs and preferences of the user. After serving the current request, the modification is discarded and the model is restored to the state when the current request is received.  

We introduce two versions of LAST algorithms with more details. The first is the cascade version, which finds the optimal modification $\Delta \boldsymbol{\theta}^*(u, \mathcal{C})$ through an iterative process as shown in Fig.~\ref{fig:LAST}~(b). Upon the arrival of a new user request, the cascade approach commences by generating a prediction using the currently deployed model and subsequently invokes the evaluation function, $E(\cdot)$, to assess the quality of this prediction. Thereafter, LAST endeavors to adjust the model parameters, $\boldsymbol{\theta}$, in an effort to enhance the evaluation score. The specific methods employed for updating may vary based on the application context and can range from straightforward gradient descent techniques to more intricate RL strategies. Following this initial modification of the model, LAST attempts to produce a revised list of recommendations. It proceeds to engage in a cycle of prediction, evaluation, and parameter updating until a predetermined stopping criterion is satisfied. This criterion can be defined as a minimal change in evaluation per iteration or a maximum number of iterations. The recommendation list produced in the final iteration is presented to the user. Although the cascade version of LAST can be highly effective, its iterative nature has the potential to lead to a substantial increase in the system's response time. This delay may render the cascade variant less suitable for online serving systems, where a rapid response is crucial.

The parallel version of LAST offers enhanced efficiency compared to its cascade counterpart, as shown in Fig.~\ref{fig:LAST}~(c). It introduces a new gradient exploration module that suggests potential model modifications. After receiving the modification suggestions, the generator attempts the suggestions, and $E(\cdot)$ estimates their effectiveness. The list with the highest evaluation is presented to the user. Algorithm.~\ref{tab:BLAST} shows a more concrete implementation. We explore two very special gradient directions: the one increasing the generating probability of the current list and its opposite. It is interesting that finding these directions does not involve the execution of $E(\cdot)$. Mirroring the principles of offline training, we enhance the probability of list generation when a high reward is obtained, and conversely, lower it when the reward is minor. The value obtained from $E(\cdot)$ is instrumental in calibrating the magnitude of the updates, but not the direction. The list generating probability $P(L)$ can be easily derived from a generative actor. The gradient is normalized with respect to the magnitude of $\boldsymbol{\theta}$. The normalized gradient and a set of manually specified step sizes are utilized to provide the model modification suggestions. The partial gradient of $E(\cdot)$ with respect to $\boldsymbol{\theta}$ is not directly used because $E(\cdot)$ may not be a function of  $\boldsymbol{\theta}$  and thus the partial gradient does not exist. Modern neural network models can have billions of parameters. It is not necessary or efficient to modify all of them for a single request. A more feasible solution is to modify only a key subset of $\boldsymbol{\theta}$.

\begin{algorithm}
    \caption{LAST, the parallel version implementation.}
    \label{tab:BLAST}
    \begin{algorithmic}[1]
    \REQUIRE a user $u$; a candidate item set $\mathcal{C}$; a deployed actor model $G(\cdot; \boldsymbol{\theta})$ parameterized by $\boldsymbol{\theta}$; a list evaluation function $E(\cdot)$; a function $P(L)$ indicating the probability of $G$ generating list $L$; a list of step size $[\eta_1, \eta_2, ...]$; a constant factor $\alpha$ for gradient normalization 
    \ENSURE a predicted optimal recommendation list $\widehat{L^*}_{LAST}$ for $u$, which is composed by items in $\mathcal{C}$
    \STATE $\hat{L} \leftarrow G(u, \mathcal{C}; \boldsymbol{\theta})$ \COMMENT{run the prediction model}
    \STATE $\boldsymbol{g}_{\boldsymbol{\theta}} \leftarrow \frac{\partial P(\hat{L})}{\partial \boldsymbol{\theta}}$ \COMMENT{calculate the partial derivative of $P$ with respect to $\boldsymbol{\theta}$}
    \STATE $\Delta \boldsymbol{\theta} \leftarrow \alpha  \frac{\left | \boldsymbol{\theta} \right |}{\left | \boldsymbol{g}_{\boldsymbol{\theta}} \right |} \boldsymbol{g}_{\boldsymbol{\theta}}$ \COMMENT{normalize the gradient}
    \FOR{$\eta$ in $[\eta_1, \eta_2, ...]$}                    
        \STATE $\hat{L}_{\eta} \leftarrow G(u, \mathcal{C}; \boldsymbol{\theta} + \eta \Delta \boldsymbol{\theta})$
        \STATE $E_{\eta} \leftarrow E(u, \hat{L}_{\eta})$
    \ENDFOR
    \STATE $\eta^*=\mathrm{argmax}_{\eta} E_{\eta}$
    \RETURN $\widehat{L^*}_{LAST}=L_{\eta^*}$
    \end{algorithmic}
\end{algorithm}

\section{Experiments}\label{sec:experiment}

\subsection{Offline Experiments}\label{sec:offline}

We conduct our offline experiments on the benchmark LibRerank\footnote{https://github.com/LibRerank-Community/LibRerank\label{LibRerank}} with the public recommendation dataset Ad2\footnote{https://tianchi.aliyun.com/dataset/56}. The source code has been released\footnote{https://anonymous.4open.science/r/LAST-BCCE}. For more details on the experimental setup, please refer to Auxiliary Materials. Our offline experiments aim to answer two key questions: (i) Does LAST outperform the latest re-ranking models? and (ii) How do hyper-parameters impact the effectiveness of LAST?

\subsubsection{Performance Analysis}\label{sec:performance analysis}
In the first offline experiment, we train re-ranking models to boost user engagement. To make a fair comparison between AE and non-AE models, we use NDCG as the evaluation function to train AE models. The non-AE methods obtain the final recommendation list by ranking candidates according to the engagement probability prediction of each item from high to low.
Table.~\ref{tab:offline_experiment1} summarizes the results. It is interesting to see that the non-AE methods, including GSF\cite{ai2019learning}, DLCM\cite{ai2018learning}, SetRank\cite{pang2020setrank}, and PRM\cite{pei2019personalized}, outperform the AE methods, i.e. EGR\cite{wang2019sequential} and CMR(Greedy)\cite{chen2023controllable}. These AE methods use a single-list strategy, which means the actor will only generate a single recommendation list, by picking the item with the largest selection probability in each step. We think it is because "ranking the items according to the predicted user engagement probabilities" is a very strong prior, it significantly reduced the modeling difficulty. RL searching algorithms do not know this prior and thus converge to a worse local optimal. When it comes to AE re-ranking models with the multi-list strategy including CMR(Sampling) and LAST, they beat other baselines by a large margin. It reflects the simple fact that more trials can always lead to better results, in probability. The margins seem surprisingly large. This is because MAP and NDCG are discrete functions and heavily emphasize the top lists.  For example, the NDCG value of the perfect list "1, 0, 0, 0, 0" is 1 and 0.63 for a close list "0, 1, 0, 0, 0". The former is nearly 59 percent higher than the latter, which appears to be exaggerated. Here 1 in the list means a relevant item and 0 means an irrelevant one. LAST performs significantly better than all other baselines, which supports the benefit of serving time adaptation. 

\begin{table}[htbp]
\caption{Offline comparison of re-ranking models. The AE re-ranking models use the NDCG metric as the evaluator. The best results is bolded and the runner-up is underlined. * indicates that the improvement over the best baseline is statistically significant ($p$-value < 0.05).}
\label{tab:offline_experiment1}
\vspace{-2ex}
\begin{tabular}{lllll}
\toprule
Method & map@5 & map@10 & ndcg@5 & ndcg@10 \\ 
\midrule
GSF            & 0.6038 & 0.6074 & 0.6838 & 0.6984 \\
DLCM           & 0.6169 & 0.6201 & 0.6948 & 0.7080 \\
SetRank        & 0.6084 & 0.6122 & 0.6878 & 0.7020 \\
PRM            & 0.6176 & 0.6209 & 0.6950 & 0.7087 \\
EGR            & 0.6010 & 0.6047 & 0.6822 & 0.6964 \\
CMR(Greedy)    & 0.6032 & 0.6067 & 0.6838 & 0.6979 \\
CMR(Sampling)  & \underline{0.6559} & \underline{0.6587} & \underline{0.7243} & \underline{0.7370} \\
LAST           & \textbf{0.6623}$^{*}$ & \textbf{0.6652}$^{*}$ & \textbf{0.7289}$^{*}$ & \textbf{0.7418}$^{*}$\\
\bottomrule
\vspace{-4ex}
\end{tabular}
\end{table}

In the second offline experiment, we train the AE re-ranking models with a pre-trained evaluator, which is the standard practice of the AE framework, leading to better online performance experimentally. In this experiment, the model structure of the evaluator is the same as in CMR, and it tries to predict the click probability of each item. We use evaluator@N to indicate the quality of the recommendation list, which represents the item-wise average click probability of the top N items. The results in Table.~\ref{tab:offline_experiment2} clearly show the advantage of re-ranking models with the multi-list strategy over the ones with the single-list strategy, and the advantage of LAST over baselines. The relative improvements appear to be more reasonable compared to the first offline experiment, which is more likely to indicate the improvement online.  

\begin{table}[htbp]
\caption{Offline comparison of re-ranking models. The AE re-ranking models use a pre-trained model evaluator to predict user engagement. The best results is bolded and the runner-up is underlined. * indicates that the improvement over the best baseline is statistically significant ($p$-value < 0.05).}
\label{tab:offline_experiment2}
\vspace{-2ex}
\begin{tabular}{lllll}
\toprule
Method & evaluator@5 & evaluator@10 \\ 
\midrule
EGR            & 0.3215 & 0.3086 \\
CMR(Greedy)    & 0.3259 & 0.3131 \\
CMR(Sampling)  & \underline{0.3395} & \underline{0.3267} \\
LAST           & \textbf{0.3438}$^{*}$ & \textbf{0.3310}$^{*}$\\
\bottomrule
\vspace{-4ex}
\end{tabular}
\end{table}

\subsubsection{Hyper-parameters Analysis}\label{sec:hyper_parameter_analysis}
\begin{figure}[htb]
     \centering
     \begin{subfigure}[l]{0.23\textwidth}
         \includegraphics[width=\textwidth]{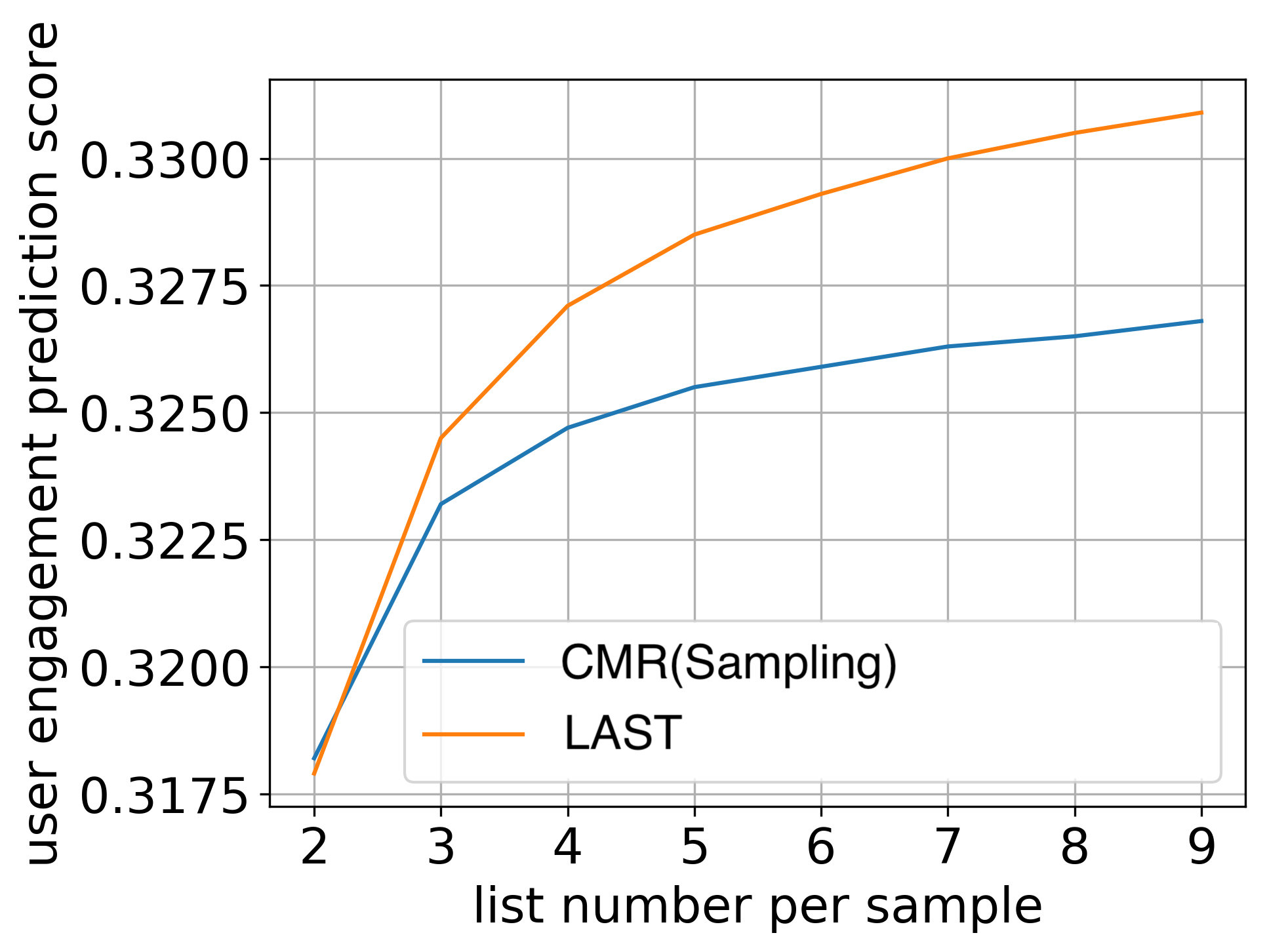}
         \label{fig:step}
     \end{subfigure}
     \hfill
     \begin{subfigure}[r]{0.23\textwidth}
         \includegraphics[width=\textwidth]{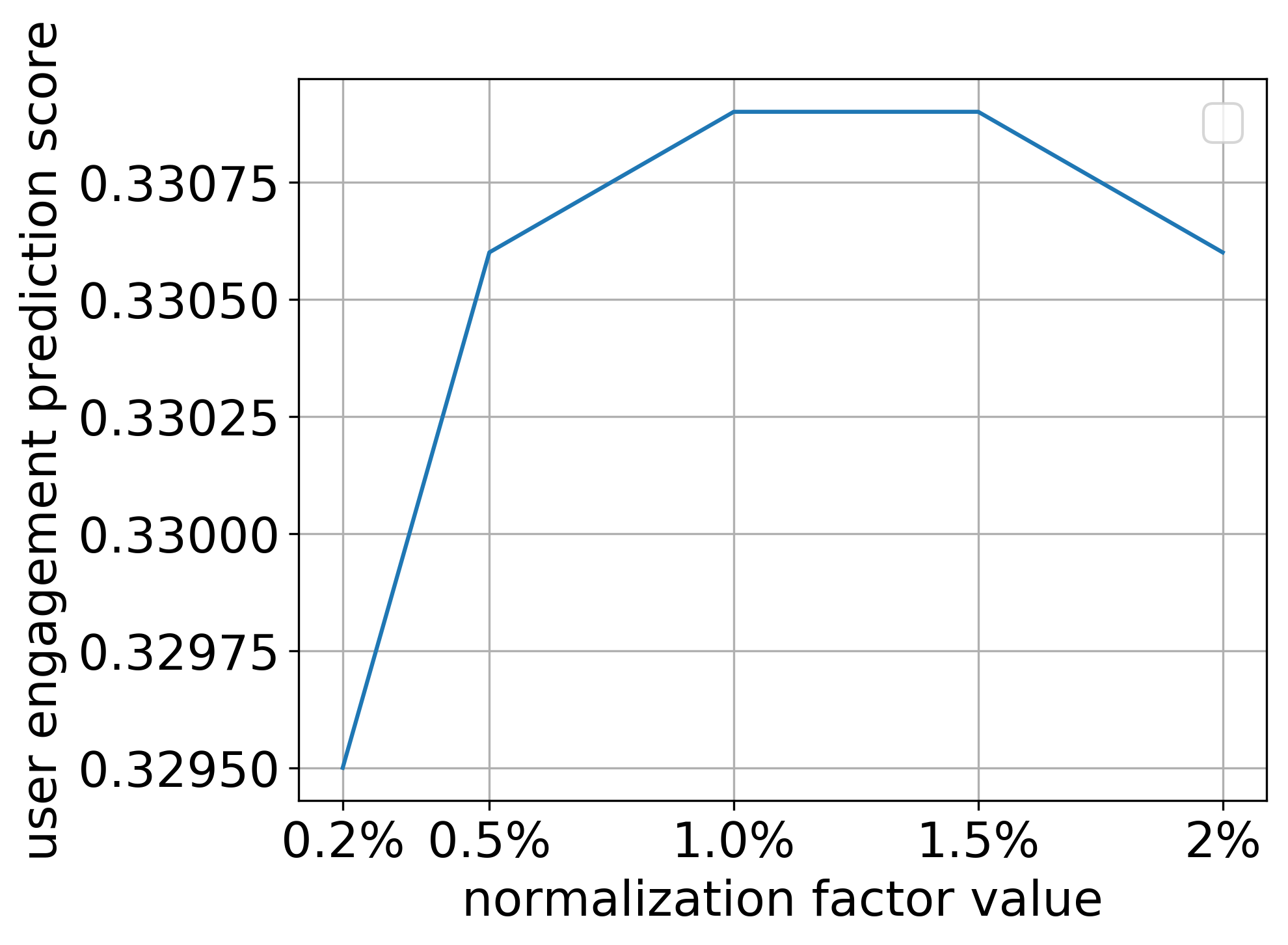}
         \label{fig:norm}
     \end{subfigure}
        \caption{The impact of hyper-parameters on LAST.}
        \label{fig:hyper_parameter}
    \vspace{-2ex}
\end{figure}

We evaluate the impact of two core hyper-parameters of LAST in the third offline experiment. Fig.~\ref{fig:hyper_parameter}(a) shows the influence of the length of the step size lists $[\eta_1, \eta_2, ...]$, where each $\eta$ generates a recommendation list. We can see that more list trials always lead to higher evaluation scores and LAST is clearly more effective than CMR(Sampling). Fig.~\ref{fig:hyper_parameter}(b) depicts the influence of the normalization factor $\alpha$. When $\alpha$ is too small, the modification is not strong enough to effectively change the recommendation lists to increase the evaluation score; when  $\alpha$ goes too big, the direction indicated by the local gradient becomes unreliable. Both lead to a suppressed LAST improvement. The performance of LAST reaches its peak when $\alpha$ is 1\%.  

\subsection{Online Experiments}\label{sec:online}
We conduct online experiments on the "Subscribe" scene in the Taobao App, a leading e-commerce platform in China. We aim to improve the purchase number per user, meanwhile maintaining the user interaction frequency, namely the click number per user. CMR(Sampling) is the baseline and LAST is the experiment method. For more details on the experimental setup, please refer to Auxiliary Materials.
The results from the online A/B testing are summarized in Table~\ref{tab:online_ab}, where the superiority of LAST is evident. By explicitly adapting its model parameters in real-time to each incoming request, LAST is able to achieve a higher conversion rate, leading to more purchases, without detracting from user engagement.

 \begin{table}[!htb]
\caption{Online A/B test results, LAST over CMR(Sampling).}
\vspace{-2ex}
\label{tab:online_ab}
\small{
\begin{tabular}{lr}
\toprule
Metric & Relative improvement \\
\midrule
click number per user & 0.08\% \\
purchase number per user & 2.08\% \\
\bottomrule
\vspace{-6ex}
\end{tabular}}
\end{table}

\section{Conclusion}
We propose a novel re-ranking model in e-commerce with a unique online learning ability. Distinct from existing methods, it can achieve effective model online learning without waiting for real user feedback, by using supervision signals provided by a surrogate model. Deliberated algorithmic designs are introduced to fulfill its full potential. It introduces request-specific modifications to maximize model adaptation to the context of each request. It discards the modifications after each recommendation to stabilize the training procedure. These designs ensure LAST can work in harmony with existing online learning systems while providing improved temporal effectiveness and incremental adaptation. Through extensive offline and online experiments, we demonstrate the strength of the new proposed model.

\bibliographystyle{ACM-Reference-Format}
\bibliography{ref}


\begin{thebibliography}{25}


\ifx \showCODEN    \undefined \def \showCODEN     #1{\unskip}     \fi
\ifx \showDOI      \undefined \def \showDOI       #1{#1}\fi
\ifx \showISBNx    \undefined \def \showISBNx     #1{\unskip}     \fi
\ifx \showISBNxiii \undefined \def \showISBNxiii  #1{\unskip}     \fi
\ifx \showISSN     \undefined \def \showISSN      #1{\unskip}     \fi
\ifx \showLCCN     \undefined \def \showLCCN      #1{\unskip}     \fi
\ifx \shownote     \undefined \def \shownote      #1{#1}          \fi
\ifx \showarticletitle \undefined \def \showarticletitle #1{#1}   \fi
\ifx \showURL      \undefined \def \showURL       {\relax}        \fi
\providecommand\bibfield[2]{#2}
\providecommand\bibinfo[2]{#2}
\providecommand\natexlab[1]{#1}
\providecommand\showeprint[2][]{arXiv:#2}

\bibitem[Ai et~al\mbox{.}(2018)]%
        {ai2018learning}
\bibfield{author}{\bibinfo{person}{Qingyao Ai}, \bibinfo{person}{Keping Bi}, \bibinfo{person}{Jiafeng Guo}, {and} \bibinfo{person}{W~Bruce Croft}.} \bibinfo{year}{2018}\natexlab{}.
\newblock \showarticletitle{Learning a deep listwise context model for ranking refinement}. In \bibinfo{booktitle}{\emph{The 41st international ACM SIGIR conference on research \& development in information retrieval}}. \bibinfo{pages}{135--144}.
\newblock


\bibitem[Ai et~al\mbox{.}(2019)]%
        {ai2019learning}
\bibfield{author}{\bibinfo{person}{Qingyao Ai}, \bibinfo{person}{Xuanhui Wang}, \bibinfo{person}{Sebastian Bruch}, \bibinfo{person}{Nadav Golbandi}, \bibinfo{person}{Michael Bendersky}, {and} \bibinfo{person}{Marc Najork}.} \bibinfo{year}{2019}\natexlab{}.
\newblock \showarticletitle{Learning groupwise multivariate scoring functions using deep neural networks}. In \bibinfo{booktitle}{\emph{Proceedings of the 2019 ACM SIGIR international conference on theory of information retrieval}}. \bibinfo{pages}{85--92}.
\newblock


\bibitem[Bello et~al\mbox{.}(2018)]%
        {bello2018seq2slate}
\bibfield{author}{\bibinfo{person}{Irwan Bello}, \bibinfo{person}{Sayali Kulkarni}, \bibinfo{person}{Sagar Jain}, \bibinfo{person}{Craig Boutilier}, \bibinfo{person}{Ed Chi}, \bibinfo{person}{Elad Eban}, \bibinfo{person}{Xiyang Luo}, \bibinfo{person}{Alan Mackey}, {and} \bibinfo{person}{Ofer Meshi}.} \bibinfo{year}{2018}\natexlab{}.
\newblock \showarticletitle{Seq2slate: Re-ranking and slate optimization with rnns}.
\newblock \bibinfo{journal}{\emph{arXiv preprint arXiv:1810.02019}} (\bibinfo{year}{2018}).
\newblock


\bibitem[Chen et~al\mbox{.}(2023)]%
        {chen2023controllable}
\bibfield{author}{\bibinfo{person}{Sirui Chen}, \bibinfo{person}{Yuan Wang}, \bibinfo{person}{Zijing Wen}, \bibinfo{person}{Zhiyu Li}, \bibinfo{person}{Changshuo Zhang}, \bibinfo{person}{Xiao Zhang}, \bibinfo{person}{Quan Lin}, \bibinfo{person}{Cheng Zhu}, {and} \bibinfo{person}{Jun Xu}.} \bibinfo{year}{2023}\natexlab{}.
\newblock \showarticletitle{Controllable Multi-Objective Re-ranking with Policy Hypernetworks}. In \bibinfo{booktitle}{\emph{Proceedings of the 29th ACM SIGKDD Conference on Knowledge Discovery and Data Mining}}. \bibinfo{pages}{3855--3864}.
\newblock


\bibitem[Gong et~al\mbox{.}(2019)]%
        {gong2019exact}
\bibfield{author}{\bibinfo{person}{Yu Gong}, \bibinfo{person}{Yu Zhu}, \bibinfo{person}{Lu Duan}, \bibinfo{person}{Qingwen Liu}, \bibinfo{person}{Ziyu Guan}, \bibinfo{person}{Fei Sun}, \bibinfo{person}{Wenwu Ou}, {and} \bibinfo{person}{Kenny~Q Zhu}.} \bibinfo{year}{2019}\natexlab{}.
\newblock \showarticletitle{Exact-k recommendation via maximal clique optimization}. In \bibinfo{booktitle}{\emph{Proceedings of the 25th ACM SIGKDD international conference on knowledge discovery \& data mining}}. \bibinfo{pages}{617--626}.
\newblock


\bibitem[Hazan(2016)]%
        {Hazan2016Introduction}
\bibfield{author}{\bibinfo{person}{Elad Hazan}.} \bibinfo{year}{2016}\natexlab{}.
\newblock \showarticletitle{Introduction to online convex optimization}.
\newblock \bibinfo{journal}{\emph{Foundations and Trends$^{\textregistered}$ in Optimization}} \bibinfo{volume}{2}, \bibinfo{number}{3-4} (\bibinfo{year}{2016}), \bibinfo{pages}{157--325}.
\newblock


\bibitem[Huzhang et~al\mbox{.}(2021)]%
        {huzhang2021aliexpress}
\bibfield{author}{\bibinfo{person}{Guangda Huzhang}, \bibinfo{person}{Zhenjia Pang}, \bibinfo{person}{Yongqing Gao}, \bibinfo{person}{Yawen Liu}, \bibinfo{person}{Weijie Shen}, \bibinfo{person}{Wen-Ji Zhou}, \bibinfo{person}{Qing Da}, \bibinfo{person}{Anxiang Zeng}, \bibinfo{person}{Han Yu}, \bibinfo{person}{Yang Yu}, {et~al\mbox{.}}} \bibinfo{year}{2021}\natexlab{}.
\newblock \showarticletitle{AliExpress Learning-To-Rank: Maximizing online model performance without going online}.
\newblock \bibinfo{journal}{\emph{IEEE Transactions on Knowledge and Data Engineering}} (\bibinfo{year}{2021}).
\newblock


\bibitem[Jiang et~al\mbox{.}(2017)]%
        {jiang2017learning}
\bibfield{author}{\bibinfo{person}{Zhengbao Jiang}, \bibinfo{person}{Ji-Rong Wen}, \bibinfo{person}{Zhicheng Dou}, \bibinfo{person}{Wayne~Xin Zhao}, \bibinfo{person}{Jian-Yun Nie}, {and} \bibinfo{person}{Ming Yue}.} \bibinfo{year}{2017}\natexlab{}.
\newblock \showarticletitle{Learning to diversify search results via subtopic attention}. In \bibinfo{booktitle}{\emph{Proceedings of the 40th international ACM SIGIR Conference on Research and Development in Information Retrieval}}. \bibinfo{pages}{545--554}.
\newblock


\bibitem[Kaelbling et~al\mbox{.}(1996)]%
        {kaelbling1996reinforcement}
\bibfield{author}{\bibinfo{person}{Leslie~Pack Kaelbling}, \bibinfo{person}{Michael~L Littman}, {and} \bibinfo{person}{Andrew~W Moore}.} \bibinfo{year}{1996}\natexlab{}.
\newblock \showarticletitle{Reinforcement learning: A survey}.
\newblock \bibinfo{journal}{\emph{Journal of artificial intelligence research}}  \bibinfo{volume}{4} (\bibinfo{year}{1996}), \bibinfo{pages}{237--285}.
\newblock


\bibitem[Karako and Manggala(2018)]%
        {karako2018using}
\bibfield{author}{\bibinfo{person}{Chen Karako} {and} \bibinfo{person}{Putra Manggala}.} \bibinfo{year}{2018}\natexlab{}.
\newblock \showarticletitle{Using image fairness representations in diversity-based re-ranking for recommendations}. In \bibinfo{booktitle}{\emph{Adjunct Publication of the 26th Conference on User Modeling, Adaptation and Personalization}}. \bibinfo{pages}{23--28}.
\newblock


\bibitem[Kotkov et~al\mbox{.}(2016)]%
        {kotkov2016survey}
\bibfield{author}{\bibinfo{person}{Denis Kotkov}, \bibinfo{person}{Shuaiqiang Wang}, {and} \bibinfo{person}{Jari Veijalainen}.} \bibinfo{year}{2016}\natexlab{}.
\newblock \showarticletitle{A survey of serendipity in recommender systems}.
\newblock \bibinfo{journal}{\emph{Knowledge-Based Systems}}  \bibinfo{volume}{111} (\bibinfo{year}{2016}), \bibinfo{pages}{180--192}.
\newblock


\bibitem[Li(2017)]%
        {li2017deep}
\bibfield{author}{\bibinfo{person}{Yuxi Li}.} \bibinfo{year}{2017}\natexlab{}.
\newblock \showarticletitle{Deep reinforcement learning: An overview}.
\newblock \bibinfo{journal}{\emph{arXiv preprint arXiv:1701.07274}} (\bibinfo{year}{2017}).
\newblock


\bibitem[Liu et~al\mbox{.}(2020)]%
        {liu2020dvgan}
\bibfield{author}{\bibinfo{person}{Jiongnan Liu}, \bibinfo{person}{Zhicheng Dou}, \bibinfo{person}{Xiaojie Wang}, \bibinfo{person}{Shuqi Lu}, {and} \bibinfo{person}{Ji-Rong Wen}.} \bibinfo{year}{2020}\natexlab{}.
\newblock \showarticletitle{DVGAN: a minimax game for search result diversification combining explicit and implicit features}. In \bibinfo{booktitle}{\emph{Proceedings of the 43rd International ACM SIGIR Conference on Research and Development in Information Retrieval}}. \bibinfo{pages}{479--488}.
\newblock


\bibitem[Oosterhuis(2021)]%
        {oosterhuis2021computationally}
\bibfield{author}{\bibinfo{person}{Harrie Oosterhuis}.} \bibinfo{year}{2021}\natexlab{}.
\newblock \showarticletitle{Computationally efficient optimization of plackett-luce ranking models for relevance and fairness}. In \bibinfo{booktitle}{\emph{Proceedings of the 44th International ACM SIGIR Conference on Research and Development in Information Retrieval}}. \bibinfo{pages}{1023--1032}.
\newblock


\bibitem[Pang et~al\mbox{.}(2020)]%
        {pang2020setrank}
\bibfield{author}{\bibinfo{person}{Liang Pang}, \bibinfo{person}{Jun Xu}, \bibinfo{person}{Qingyao Ai}, \bibinfo{person}{Yanyan Lan}, \bibinfo{person}{Xueqi Cheng}, {and} \bibinfo{person}{Jirong Wen}.} \bibinfo{year}{2020}\natexlab{}.
\newblock \showarticletitle{Setrank: Learning a permutation-invariant ranking model for information retrieval}. In \bibinfo{booktitle}{\emph{Proceedings of the 43rd International ACM SIGIR Conference on Research and Development in Information Retrieval}}. \bibinfo{pages}{499--508}.
\newblock


\bibitem[Pei et~al\mbox{.}(2019)]%
        {pei2019personalized}
\bibfield{author}{\bibinfo{person}{Changhua Pei}, \bibinfo{person}{Yi Zhang}, \bibinfo{person}{Yongfeng Zhang}, \bibinfo{person}{Fei Sun}, \bibinfo{person}{Xiao Lin}, \bibinfo{person}{Hanxiao Sun}, \bibinfo{person}{Jian Wu}, \bibinfo{person}{Peng Jiang}, \bibinfo{person}{Junfeng Ge}, \bibinfo{person}{Wenwu Ou}, {et~al\mbox{.}}} \bibinfo{year}{2019}\natexlab{}.
\newblock \showarticletitle{Personalized re-ranking for recommendation}. In \bibinfo{booktitle}{\emph{Proceedings of the 13th ACM conference on recommender systems}}. \bibinfo{pages}{3--11}.
\newblock


\bibitem[Sahoo et~al\mbox{.}(2018)]%
        {Sahoo2018Online}
\bibfield{author}{\bibinfo{person}{Doyen Sahoo}, \bibinfo{person}{Quang Pham}, \bibinfo{person}{Jing Lu}, {and} \bibinfo{person}{Steven C.~H. Hoi}.} \bibinfo{year}{2018}\natexlab{}.
\newblock \showarticletitle{Online Deep Learning: {L}earning Deep Neural Networks on the Fly}. In \bibinfo{booktitle}{\emph{Proceedings of the 27th International Joint Conference on Artificial Intelligence}}. \bibinfo{publisher}{International Joint Conferences on Artificial Intelligence Organization}, \bibinfo{pages}{2660--2666}.
\newblock


\bibitem[Vaswani et~al\mbox{.}(2017)]%
        {vaswani2017attention}
\bibfield{author}{\bibinfo{person}{Ashish Vaswani}, \bibinfo{person}{Noam Shazeer}, \bibinfo{person}{Niki Parmar}, \bibinfo{person}{Jakob Uszkoreit}, \bibinfo{person}{Llion Jones}, \bibinfo{person}{Aidan~N Gomez}, \bibinfo{person}{{\L}ukasz Kaiser}, {and} \bibinfo{person}{Illia Polosukhin}.} \bibinfo{year}{2017}\natexlab{}.
\newblock \showarticletitle{Attention is all you need}.
\newblock \bibinfo{journal}{\emph{Advances in neural information processing systems}}  \bibinfo{volume}{30} (\bibinfo{year}{2017}).
\newblock


\bibitem[Wang et~al\mbox{.}(2019)]%
        {wang2019sequential}
\bibfield{author}{\bibinfo{person}{Fan Wang}, \bibinfo{person}{Xiaomin Fang}, \bibinfo{person}{Lihang Liu}, \bibinfo{person}{Yaxue Chen}, \bibinfo{person}{Jiucheng Tao}, \bibinfo{person}{Zhiming Peng}, \bibinfo{person}{Cihang Jin}, {and} \bibinfo{person}{Hao Tian}.} \bibinfo{year}{2019}\natexlab{}.
\newblock \showarticletitle{Sequential evaluation and generation framework for combinatorial recommender system}.
\newblock \bibinfo{journal}{\emph{arXiv preprint arXiv:1902.00245}} (\bibinfo{year}{2019}).
\newblock


\bibitem[Wiering and Van~Otterlo(2012)]%
        {wiering2012reinforcement}
\bibfield{author}{\bibinfo{person}{Marco~A Wiering} {and} \bibinfo{person}{Martijn Van~Otterlo}.} \bibinfo{year}{2012}\natexlab{}.
\newblock \showarticletitle{Reinforcement learning}.
\newblock \bibinfo{journal}{\emph{Adaptation, learning, and optimization}} \bibinfo{volume}{12}, \bibinfo{number}{3} (\bibinfo{year}{2012}), \bibinfo{pages}{729}.
\newblock


\bibitem[Xia et~al\mbox{.}(2016)]%
        {xia2016modeling}
\bibfield{author}{\bibinfo{person}{Long Xia}, \bibinfo{person}{Jun Xu}, \bibinfo{person}{Yanyan Lan}, \bibinfo{person}{Jiafeng Guo}, {and} \bibinfo{person}{Xueqi Cheng}.} \bibinfo{year}{2016}\natexlab{}.
\newblock \showarticletitle{Modeling document novelty with neural tensor network for search result diversification}. In \bibinfo{booktitle}{\emph{Proceedings of the 39th International ACM SIGIR conference on Research and Development in Information Retrieval}}. \bibinfo{pages}{395--404}.
\newblock


\bibitem[Xia et~al\mbox{.}(2017)]%
        {xia2017adapting}
\bibfield{author}{\bibinfo{person}{Long Xia}, \bibinfo{person}{Jun Xu}, \bibinfo{person}{Yanyan Lan}, \bibinfo{person}{Jiafeng Guo}, \bibinfo{person}{Wei Zeng}, {and} \bibinfo{person}{Xueqi Cheng}.} \bibinfo{year}{2017}\natexlab{}.
\newblock \showarticletitle{Adapting Markov decision process for search result diversification}. In \bibinfo{booktitle}{\emph{Proceedings of the 40th international ACM SIGIR conference on research and development in information retrieval}}. \bibinfo{pages}{535--544}.
\newblock


\bibitem[Zhang et~al\mbox{.}(2019)]%
        {Zhang2019Survey}
\bibfield{author}{\bibinfo{person}{Xiao Zhang}, \bibinfo{person}{Yun Liao}, {and} \bibinfo{person}{Shizhong Liao}.} \bibinfo{year}{2019}\natexlab{}.
\newblock \showarticletitle{A survey on online kernel selection for online kernel learning}.
\newblock \bibinfo{journal}{\emph{WIREs Data Mining and Knowledge Discovery}} \bibinfo{volume}{9}, \bibinfo{number}{2} (\bibinfo{year}{2019}), \bibinfo{pages}{e1295}.
\newblock
\urldef\tempurl%
\url{https://doi.org/10.1002/widm.1295}
\showDOI{\tempurl}


\bibitem[Zhu et~al\mbox{.}(2021)]%
        {zhu2021fairness}
\bibfield{author}{\bibinfo{person}{Ziwei Zhu}, \bibinfo{person}{Jingu Kim}, \bibinfo{person}{Trung Nguyen}, \bibinfo{person}{Aish Fenton}, {and} \bibinfo{person}{James Caverlee}.} \bibinfo{year}{2021}\natexlab{}.
\newblock \showarticletitle{Fairness among new items in cold start recommender systems}. In \bibinfo{booktitle}{\emph{Proceedings of the 44th International ACM SIGIR Conference on Research and Development in Information Retrieval}}. \bibinfo{pages}{767--776}.
\newblock


\bibitem[Zhuang et~al\mbox{.}(2018)]%
        {zhuang2018globally}
\bibfield{author}{\bibinfo{person}{Tao Zhuang}, \bibinfo{person}{Wenwu Ou}, {and} \bibinfo{person}{Zhirong Wang}.} \bibinfo{year}{2018}\natexlab{}.
\newblock \showarticletitle{Globally optimized mutual influence aware ranking in e-commerce search}.
\newblock \bibinfo{journal}{\emph{arXiv preprint arXiv:1805.08524}} (\bibinfo{year}{2018}).
\newblock


\end{thebibliography}

\newpage
\section{Auxiliary Material}
\subsection{Offline Experiment Setup}
The original Ad dataset records 1 million users and 26 million ad display/click logs, with 8 user profiles (e.g., id, age, and occupation), 6 item features (e.g., id, campaign, and brand). LibRerank transformed the records of each user into ranking lists according to the timestamp of the user browsing the advertisement. Items that have been interacted with within 5 minutes are sliced into a list. The final Ad dataset contains 349,404 items and 483,049 lists. 

We chose a wide range of representative and state-of-the-art re-ranking methods as baselines.
\begin{itemize}
    \item GSF \cite{ai2019learning}: This method explicitly divides item candidates into overlapping groups of size 2 or 3 and introduces a group-based scoring function to represent items in each group. 
    \item DLCM \cite{ai2018learning}: Deep Listwise Context Model uses a Gated Recurrent Unit (GRU) to capture the global context of a recommendation.
    \item PRM\cite{pei2019personalized}: The Personalized Re-ranking Model uses transformer blocks \cite{vaswani2017attention} to capture the mutual influence between item candidates.
    \item SetRank \cite{pang2020setrank}: SetRank is a Bayesian approach for collaborative ranking, which aims at maximizing the posterior probability of set-wise preference comparisons. 
   \item{EGR}~\cite{wang2019sequential}: A re-ranking model uses an evaluator-generator framework that includes a generator responsible for generating feasible permutations, and an evaluator that assesses the list-wise utility of each permutation. The framework is also known as the actor-evaluator. The two differ only in terminology.
   \item{CMR(Greedy)}~\cite{chen2023controllable}:  An enhanced version of EGR with more powerful model structures. The actor works in a greedy manner, which means the actor always picks the item with the largest selection probability in each step and generates only one recommendation list for each request. 
   \item{CMR(Sampling)}: The model structure is the same as above. The actor generates multiple recommendation lists using Thompson Sampling and the list with the highest evaluation score is presented to a user. The parameters of the actor remain the same during serving time. It is the re-ranking model that serves the main traffic of our industrial applied recommendation scene. 
   \item{LAST}: The model structure is the same as above. The actor generates multiple recommendation lists using the parallel version of LAST. The list with the highest evaluation score is presented to a user.
\end{itemize}

\subsubsection{Offline Experiment 1}
The non-AE re-ranking models, including GSF, DLCM, SetRank, and PRM, are trained with item-level user feedback labels. In other words, they try to predict user engagement of each item, considering the candidate context. The final recommendation list is obtained by ranking candidates according to the engagement probability prediction from high to low. To make a fair comparison between AE and non-AE models, in this experiment, we use NDCG as the evaluation function to train AE models. Evaluation based on NDCG uses a hidden assumption that the user feedback on each item stays the same after re-ranking the lists. It is not likely to hold in the perspective of re-ranking modeling. However, it is an unavoidable assumption in non-AE re-ranking models and is widely used in related works. 

\subsection{Online Experiment Setup}
We conduct online experiments on the "Subscribe" scene in the Taobao App, a leading e-commerce platform in China. Its main entrance is the "Subscribe" button on top of Taobao's main landing page and it is a stream of various elements including items, posters, videos, etc. 2,003,565 users were involved in the 7-day long experiments. 
In the online experiment, we aim to improve the purchase number per user, meanwhile maintaining the user interaction frequency, indicated by the click number per user. CMR(Sampling) is the baseline method, which serves the main traffic of our online scene and LAST is the experiment method. The main difference between the two is that LAST uses online request-adaptive modification. Other aspects, such as the model structure, the offline training process, and the exploration number of lists during online serving are the same between the two. We do not have classic online learning methods deployed at this moment because of engineering limitations. LAST learning functionality can be implemented as a normal model module and does not need any extra engineering support. We leave the combination of the two in further study. 

\renewcommand{\thefootnote}{\arabic{footnote}} 

\end{document}